\documentclass{amsart}
\usepackage{amsthm}
\usepackage{lmodern}
\usepackage[english]{babel}
\usepackage{amsmath}
\usepackage{amssymb}
\usepackage{mathptmx} 
\usepackage{color}
\usepackage{graphicx}
\usepackage{caption}
\usepackage{subcaption}
\usepackage{float}
\usepackage[all,cmtip]{xy}
\usepackage{bm}
\usepackage{enumerate}
\usepackage{pinlabel}
\newtheorem{theorem}{\bf Theorem}[section]
\newtheorem{lemma}[theorem]{\bf Lemma}

\newtheorem{corollary}[theorem]{\bf Corollary}
\newtheorem{proposition}[theorem]{\bf Proposition}
\newtheorem{remark}[theorem]{\bf Remark}

\newcommand{\rme}{\mathrm{e}}
\newcommand{\rmi}{\mathrm{i}}
\newcommand{\rmd}{\mathrm{d}}

\newcommand{\defeq}{\mathrel{\mathop:}=}

\begin{document}

\title{Stable knots and links in electromagnetic fields}

\author{
Benjamin Bode}
\date{}

\address{Instituto de Ciencias Matemáticas, Consejo Superior de Investigaciones Científicas, 28049 Madrid, Spain}
\email{ben.bode.2013@my.bristol.ac.uk}




\maketitle
\begin{abstract}
In null electromagnetic fields the electric and the magnetic field lines evolve like unbreakable elastic filaments in a fluid flow. In particular, their topology is preserved for all time. We prove that for every link $L$ there is such an electromagnetic field that satisfies Maxwell's equations in free space and that has closed electric and magnetic field lines in the shape of $L$ for all time.
\end{abstract}
%

\section{Introduction}\label{sec:intro}
Knotted structures appear in physical fields in a wide range of areas of theoretical physics; in liquid crystals \cite{km:2016topology, ma:2014knotted, ma:2016global}, optical fields \cite{mark}, Bose-Einstein condensates \cite{bec}, fluid flows \cite{daniel1, daniel2}, the Skyrme-Faddeev model \cite{sutcliffe} and several others.

Mathematical constructions of initially knotted configurations in physical fields make experiments and numerical simulations possible. However, the knot typically changes or disappears as the field evolves with time as prescribed by some differential equation or energy functional. There are some results regarding the existence of stationary solutions of the harmonic oscillator and the hydrogen atom \cite{danielberry, daniel:coulomb}, and the existence of solutions to certain Schrödinger equations that describe any prescribed time evolution of a knot \cite{danielschr}. In particular, this implies the existence of solutions that contain a given knot for all time, i.e., the knot is \textit{stable} or \textit{robust}. However, more general (i.e., regarding more general differential equations) explicit analytic constructions of such solutions are not known. 

In the case of electromagnetic fields and Maxwell's equations, the first knotted solution was found by Ra\~nada \cite{ranada}. His field contains closed magnetic and electric field lines that form the Hopf link for all time. Using methods from \cite{bode:2016polynomial} and \cite{weaving} we can algorithmically construct for any given link $L$ a vector field $\mathbf{B}:\mathbb{R}^3\to\mathbb{R}^3$ that has a set of closed field lines in the shape of $L$ and that can be taken as an initial configuration of the magnetic part of an electromanetic field, say at time $t=0$. However, these links cannot be expected to be stable, since they usually undergo reconnection events as time progresses and the field evolves according to Maxwell's equations, or they disappear altogether. Necessary and sufficient conditions for the stability of knotted field lines are known \cite{kedia2}, but so far only the family of torus links has been constructed and thereby been proven to arise as stable knotted field lines in electromagnetism.

In \cite{kedia} Kedia et al. offer a construction of null electromagnetic fields with stable torus links as closed electric and magnetic field lines using an approach developed by Bateman \cite{bateman}. In this article we prove that their construction can be extended to any link type, implying the following result:

\begin{theorem}
\label{thm:main}
For every $n$-component link $L=L_1\cup L_2\cup\cdots\cup L_n$ and every subset $I\subset\{1,2,\ldots,n\}$ there is an electromagnetic field $\mathbf{F}$ that satisfies Maxwell's equations in free space and that has a set of closed field lines (electric or magnetic) ambient isotopic to $L$ for all time, with closed electric field lines that are ambient isotopic to $\bigcup_{i\in I}L_i$ for all time and closed magnetic field lines that are ambient isotopic to $\bigcup_{i\notin I}L_i$ for all time.
\end{theorem}

This shows not only that every pair of links $L_1$ and $L_2$ can arise as a set of robust closed electric and magnetic field lines, respectively, but also that any linking between the components of $L_1$ and $L_2$ can be realised.

We would like to point out that the subset $I$ of the set of components of $L$ does not need to be non-empty or proper for the theorem to hold. As a special case, we may choose $L$ and $I$ such that $\bigcup_{i\in I}L_i$ and $\bigcup_{i\notin I}L_i$ are ambient isotopic, which shows the following generalisation of the results in \cite{kedia}.
\begin{corollary}
For any link $L$ there is an electromagnetic field $\mathbf{F}$ that satisfies Maxwell's equations in free space and whose electric and magnetic field both have a set of closed field lines ambient isotopic to $L$ for all time.
\end{corollary}


The proof of the theorem relies on the existence of certain holomorphic functions, whose explicit construction eludes us at this moment. As a consequence, Theorem \ref{thm:main} guarantees the existence of the knotted fields, but does not allow us to provide any new examples beyond the torus link family.

The closed field lines at time $t=0$ turn out to be projections into $\mathbb{R}^3$ of real analytic Legendrian links with respect to the standard contact structure in $S^3$. This family of links has been studied by Rudolph in the context of holomorphic functions as totally tangential $\mathbb{C}$-links \cite{rudolphtt, rudolphtt2}.

The remainder of the article is structured as follows. In Section \ref{sec:background} we review some key mathematical concepts, in particular Bateman's construction of null electromagnetic fields and knots and their role in contact geometry. Section \ref{sec:leg} summarises some observations that relate the problem of constructing knotted field lines to a problem on holomorphic extendability of certain functions. The proof of Theorem \ref{thm:main} can be found in Section \ref{sec:proof}, where we use results by Rudolph, Burns and Stout to show that the functions in question can in fact be extended to holomorphic functions. In Section \ref{sec:disc} we offer a brief discussion of our result and some properties of the resulting electromagnetic fields.

\ \\

\textbf{Acknowledgements:} The author is grateful to Mark Dennis, Daniel Peralta-Salas and Vera Vertesi for helpful discussions. The author was supported by JSPS KAKENHI Grant Number JP18F18751 and a JSPS Postdoctoral Fellowship as JSPS International Research Fellow.

\section{Mathematical background}
\label{sec:background}

\subsection{Knots and links}
\label{sec:knots}
For $m\in\mathbb{N}$ we write $S^{2m-1}$ for the $(2m-1)$-sphere of unit radius: 
\begin{equation}
S^{2m-1}=\{(z_1,z_2,\ldots,z_m)\in\mathbb{C}^2:\sum_{i=1}^m |z_i|^2=1\}.
\end{equation} 
Via stereographic projection we have $S^3\cong \mathbb{R}^3\cup\{\infty\}$. A \textit{link} with $n$ components in a 3-manifold $M$ is (the image of) a smooth embedding of $n$ circles $S^1\sqcup S^1\sqcup\ldots\sqcup S^1$ in $M$. A link with only one component is called a \textit{knot}. The only 3-manifolds that are relevant for this article are $M=S^3$ and $M=\mathbb{R}^3$.

Knots and links are studied up to ambient isotopy or, equivalently, smooth isotopy, that is, two links are considered equivalent if one can be smoothly deformed into the other without any cutting or gluing. This defines an equivalence relation on the set of all links and we refer to the equivalence class of a link $L$ as its \textit{link type} or, in the case of a knot, as its \textit{knot type}. It is very common to be somewhat lax with the distinction between the concept of a link and its link type. When there is no risk of confusion we will for example refer to \textit{a link $L$} even though we really mean the link type, i.e., the equivalence class, represented by $L$.

One special family of links/link types is the family of torus links $T_{p,q}$ and the equivalence classes that they represent. It consists of all links that can be drawn on the surface of an unknotted torus $\mathbb{T}=S^1\times S^1$ in $\mathbb{R}^3$ or $S^3$ and they are characterised by two integers $p$ and $q$, the number of times the link winds around each $S^1$. This definition leaves an ambiguity regarding the sign of $p$ and $q$, i.e., which direction is considered as positive wrapping around the meridian and the longitude. This ambiguity is removed by the standard convention to choose 
\begin{equation}
(\rho \rme^{\rmi q\varphi}, \sqrt{1-\rho^2}\rme^{\rmi p\varphi})
\end{equation}
as a parametrisation of the $(p,q)$-torus knot in the unit 3-sphere $S^3\subset\mathbb{C}^2$ with $p,q>0$, where the parameter $\varphi$ ranges from 0 to $2\pi$ and $\rho$ is the solution to $\rho^{|p|}=\sqrt{1-\rho^2}^{|q|}$. It follows that for positive $p$ and $q$ the complex curve $z_1^p-z_2^q=0$ intersects $S^3$ in the $(p,q)$-torus knot $T_{p,q}$ \cite{milnor}.

Knot theory is now a vast and quickly developing area of mathematics with many connections to biology, chemistry and physics. For a more extensive introduction we refer the interested reader to the standard references \cite{adams, rolfsen}. The role that knots play in physics is discussed in more detail in \cite{atiyah, kauffman}.

\subsection{Bateman's construction}
\label{sec:bateman}

Our exposition of Bateman's work follows the relevant sections in \cite{kedia}. In electromagnetic fields that are null for all time the electric and magnetic field lines evolve like unbreakable elastic in an ideal fluid flow. They are dragged in the direction of the Poynting vector field with the speed of light \cite{irvine, kedia2}. This means that the link types of any closed field lines remain unchanged for all time. In the following we represent a time-dependent electromagnetic field by its Riemann-Silberstein vector $\mathbf{F}=\mathbf{E}+\rmi \mathbf{B}$, where $\mathbf{E}$ and $\mathbf{B}$ are time-dependent real vector fields on $\mathbb{R}^3$, representing the electric and magnetic part of $\mathbf{F}$, respectively.

It was shown in \cite{kedia2} that the nullness condition
\begin{equation}
\mathbf{E}\cdot\mathbf{B}=0,\qquad \mathbf{E}\cdot\mathbf{E}-\mathbf{B}\cdot\mathbf{B}=0, \qquad\text{for all }t\in\mathbb{R}
\end{equation}
is equivalent to $\mathbf{F}$ being both null and shear-free at $t=0$, that is,
\begin{equation}
(\mathbf{E}\cdot\mathbf{B})|_{t=0}=0,\qquad (\mathbf{E}\cdot\mathbf{E}-\mathbf{B}\cdot\mathbf{B})|_{t=0}=0,
\end{equation}
and
\begin{align}
((E^i E^j-B^i B^j)\partial_j V_i)|_{t=0}&=0,\nonumber\\
((E^i B^j+E^j B^i)\partial_j V_i)|_{t=0}&=0,
\end{align}
where $\mathbf{V}=\mathbf{E}\times\mathbf{B}/|\mathbf{E}\times\mathbf{B}|$ is the normalised Poynting field and the indices $i,j=1,2,3$ enumerate the components of the fields $\mathbf{E}=(E_1,E_2,E_3)$, $\mathbf{B}=(B_1,B_2,B_3)$ and $\mathbf{V}=(V_1,V_2,V_3)$.

It is worth pointing out that the Poynting vector field $\mathbf{V}$ of a null field satisfies the Euler equation for a pressure-less flow:
\begin{equation}
\partial_t \mathbf{V}+(\mathbf{V}\cdot\nabla)\mathbf{V}=0.
\end{equation}
More analogies between null light fields and pressure-less Euler flows are summarised in \cite{kedia2}.

The transport of field lines by the Poynting field of a null electromagnetic field was made precise in \cite{kedia2}. We write $W=\frac{1}{2}(\mathbf{E}\cdot\mathbf{E}+\mathbf{B}\cdot\mathbf{B})$ for the electromagnetic density. The normalised Poynting vector field $\mathbf{V}$ transports (where it is defined) $\mathbf{E}/W$ and $\mathbf{B}/W$. In the following construction $V$ can be defined everywhere and since, $\partial_t W + \nabla\cdot(W\mathbf{V})=0$, the nodal set of $W$ is also transported by $\mathbf{V}$. This implies that if $L_1$ is a link formed by closed electric field lines at time $t=0$ and $L_2$ is a link formed by closed magnetic field lines of such an electromagnetic field at $t=0$ (and in particular $W\neq 0$ on $L_1$ and $L_2$), then their time evolution according to Maxwell's equations does not only preserve the link types of $L_1$ and $L_2$, but also the way in which they are linked, i.e., the link type of $L_1\cup L_2$.

Bateman discovered a construction of null electromagnetic fields \cite{bateman}, which guarantees the stability of links and goes as follows. Take two functions $\alpha, \beta:\mathbb{R}\times\mathbb{R}^3\to\mathbb{C}$ that satisfy
\begin{equation}
\label{eq:1}
\nabla \alpha\times\nabla\beta=\rmi (\partial_t\alpha\nabla\beta-\partial_t\beta\nabla\alpha),
\end{equation}
where $\nabla$ denotes the gradient with respect to the three spatial variables.

Then for any pair of holomorphic functions $f,g:\mathbb{C}^2\to\mathbb{C}$ the field defined by
\begin{equation}
\mathbf{F}=\mathbf{E}+\rmi \mathbf{B}=\nabla f(\alpha,\beta)\times\nabla g(\alpha,\beta)
\end{equation}
satisfies Maxwell's equations and is null for all time. The field $\mathbf{F}$ can be rewritten as
\begin{equation}
\mathbf{F}=h(\alpha, \beta)\nabla\alpha\times\nabla\beta,
\end{equation}
where $h=\partial_{z_1} f\partial_{z_2} g-\partial_{z_2} f\partial_{z_1} g$ and $(z_1,z_2)$ are the coordinates in $\mathbb{C}^2$. Since $f$ and $g$ are arbitrary holomorphic functions, we obtain a null field for any holomorphic function $h:\mathbb{C}^2\to\mathbb{C}$.

Kedia et al. used Bateman's construction to find concrete examples of electromagnetic fields with knotted electric and magnetic field lines \cite{kedia}. In their work both the electric and the magnetic field lines take the shape of torus knots and links. They consider
\begin{align}
\label{eq:stereo3}
\alpha&=\frac{x^2+y^2+z^2-t^2-1+2\rmi z}{x^2+y^2+z^2-(t-\rmi)^2},\nonumber\\
\beta&=\frac{2(x-\rmi y)}{x^2+y^2+z^2-(t-\rmi)^2},
\end{align}
where $x$, $y$ and $z$ are the three spatial coordinates and $t$ represents time. It is a straightforward calculation to check that $\alpha$ and $\beta$ satisfy Equation (\ref{eq:1}). Note that for any value of $t=t_*$, the function $(\alpha,\beta)|_{t=t_*}:\mathbb{R}^3\to\mathbb{C}^2$ gives a diffeomorphism from $\mathbb{R}^3\cup\{\infty\}$ to $S^{3}\subset \mathbb{C}^2$. 

The construction of stable knots and links in electromagnetic fields therefore comes down to finding holomorphic functions $f$ and $g$, or equivalently one holomorphic function $h$. Since the image of $(\alpha, \beta)$ is $S^3$, it is not necessary for these functions to be holomorphic (or even defined) on all of $\mathbb{C}^2$. It suffices to find functions that are holomorphic on an open neighbourhood of $S^3$ in $\mathbb{C}^2$.

Kedia et al. find that for $f(z_1,z_2)=z_1^p$ and $g(z_1,z_2)=z_2^q$ the resulting electric and magnetic fields both contain field lines that form the $(p,q)$-torus link $T_{p,q}$. Hence there is a construction of flow lines in the shape of torus links that are stable for all time.

\begin{remark}
\label{remark}
It was wrongly stated in \cite{kedia} and \cite{quasi} that for $t=0$ the map $(\alpha,\beta)$ in Equation (\ref{eq:stereo3}) is the inverse of the standard stereographic projection. In fact, the inverse of the standard stereographic projection is given by $(u,v):\mathbb{R}^3\to S^3$,
\begin{align}
\label{eq:stereo4}
u&=\frac{x^2+y^2+z^2-1+2\rmi z}{x^2+y^2+z^2+1},\nonumber\\
v&=\frac{2(x+\rmi y)}{x^2+y^2+z^2+1},
\end{align}
so that $(\alpha,\beta)|_{t=0}$ is actually the inverse of the standard stereographic projection followed by a mirror reflection that sends $\text{Im}(z_2)$ to $-\text{Im}(z_2)$ or equivalently it is a mirror reflection in $\mathbb{R}^3$ along the $y=0$-plane followed by the inverse of the standard stereographic projection.

Kedia et al.'s choice of $f$ and $g$ was (in their own words) `guided' by the hypersurface $z_1^p\pm z_2^q=0$. Complex hypersurfaces like this and their singularities have been extensively studied by Milnor and others \cite{brauner, milnor} and it is well-known that the hypersurface intersects $S^3$ in the $(p,q)$-torus knot $T_{p,q}$. Even though this made the choice of $f$ and $g$ somewhat intuitive (at least for Kedia et al.), there seems to be no obvious relation between the hypersurface and the electromagnetic field that would enable us to generalise their approach. Since their fields contain the links $T_{p,q}$ in $\mathbb{R}^3$, the corresponding curves on $S^3$ are actually the mirror image $T_{p,-q}$. Therefore, it seems more plausible that (if there is a connection to complex hypersurfaces at all) the relevant complex curve is $z_1^pz_2^q-1=0$, which intersects a 3-sphere of an appropriate radius in $T_{p,-q}$ \cite{rudolphtt}. However, in contrast to Milnor's hypersurfaces, this intersection is \textit{totally tangential}, i.e., at every point of intersection the tangent plane of the hypersurface lies in the tangent space of the 3-sphere. This is an interesting property that plays an important role in the generalisation of the construction to arbitrarily complex link types in the following sections. 
\end{remark}

\subsection{Contact structures and Legendrian links}
\label{sec:contact}

A \textit{contact structure} on a 3-manifold $M$ is a smooth, completely non-integrable plane distribution $\xi\subset TM$ in the tangent bundle of $M$. It can be given as the kernel of a differential 1-form, a \textit{contact form} $\alpha$, for which the non-integrability condition reads
\begin{equation}
\alpha\wedge\rmd\alpha\neq 0.
\end{equation}
It is a convention to denote contact forms by $\alpha$. This should not be confused with the first component of the map $(\alpha,\beta)$ in Equation (\ref{eq:stereo3}). Within this subsection $\alpha$ refers to a contact form, in all other sections it refers to Equation (\ref{eq:stereo3}). The choice of $\alpha$ for a given $\xi$ is not unique, but the non-integrability property is independent of this choice.

In other words, for every point $p\in M$ we have a plane (a 2-dimensional linear subspace) $\xi_p$ in the tangent space $T_p(M)$ given by $\xi_p=\text{ker}_p\alpha$, which is the kernel of $\alpha$ when $\alpha$ is regarded as a map $T_pM\to\mathbb{R}$. The non-integrability condition ensures that there is a certain twisting of these planes throughout $M$. We call the pair of manifold $M$ and contact structure $\xi$ a \textit{contact manifold} $(M,\xi)$.

The \textit{standard contact structure} $\xi_0$ on $S^3$ is given by the contact form
\begin{equation}
\alpha_0=\sum_{j=1}^2 (x_j\rmd y_j-y_j\rmd x_j),
\end{equation}
where we write the complex coordinates $(z_1,z_2)$ of $\mathbb{C}^2$ in terms of their real and imaginary parts: $z_j=x_j+\rmi y_j$.

There are two interesting geometric interpretations of the standard contact structure $\xi_0$. Firstly, the planes are precisely the normals to the fibers of the Hopf fibration $S^3\to S^2$. Secondly, the planes are precisely the complex tangent lines to $S^3$.

A link $L$ in a contact manifold $(M,\xi)$ is called a \textit{Legendrian link} with respect to the contact structure $\xi$, if it is everywhere tangent to the contact planes, i.e., $T_pL\subset \xi_p$. It is known that every link type in $S^3$ has representatives that are Legendrian. In other words, for every link $L$ in $S^3$ there is a Legendrian link with respect to the standard contact structure on $S^3$ that is ambient isotopic to $L$.

More details on contact geometry and the connection to knot theory can be found in \cite{etnyre, geiges}.

\section{Legendrian field lines}
\label{sec:leg}

In this section we would like to point out some observations on Bateman's construction. Bateman's construction turns the problem of constructing null fields with knotted field lines into a problem of finding appropriate holomorphic functions $h:\mathbb{C}^2\to\mathbb{C}$. Our observations turn this into the question whether for a given Legendrian link $L$ with respect to the standard contact structure on $S^3$ a certain function defined on $L$ admits a holomorphic extension.

\begin{lemma}
\label{revlemma}
Let $h:\mathbb{C}^2\to\mathbb{C}$ be a function that is holomorphic on an open neighbourhood of $S^3$ and let $\mathbf{F}=h(\alpha,\beta)\nabla\alpha\times\nabla\beta$ be the corresponding electromagnetic field with $(\alpha,\beta)$ as in Equation (\ref{eq:stereo3}). Suppose $L$ is a set of closed magnetic field lines or a set of closed electric field lines of $\mathbf{F}$ at time $t=0$. Then $(\alpha,\beta)|_{t=0}(L)$ is a Legendrian link with respect to the standard contact structure on $S^3$.
\end{lemma}
\textit{Proof}: It is known that all fields that are constructed with the same choice of $(\alpha,\beta)$ have the same Poynting field, independent of $h$. For $(\alpha,\beta)$ as in Equation (\ref{eq:stereo3}) with $t=0$ its pushforward by $(\alpha,\beta)|_{t=0}$ is tangent to the fibers of the Hopf fibration. By the definition of the Poynting field, the electric and magnetic field are orthogonal to the Poynting field and it is a simple calculation that their pushforwards by $(\alpha,\beta)|_{t=0}$ are orthogonal as well. Therefore, they must be normal to the fibers of the Hopf fibration. Hence the pushforward of all electric and magnetic field lines by $(\alpha,\beta)$ are tangent to the standard contact structure on $S^3$. In particular, any closed electric or magnetic field line is a Legendrian link with respect to the standard contact structure. \qed

A more general statement of Lemma \ref{revlemma} is proven in \cite{quasi}. It turns out that $(\alpha,\beta)$ define a contact structure for each value of $t$, where time evolution is given by a 1-parameter family of contactomorphisms, and all sets of closed flow lines at a fixed moment in time are (the images in $\mathbb{R}^3$ of) Legendrian links with respect to the corresponding contact structure.

Lemma \ref{revlemma} tells us that (the projection of) closed field lines form Legendrian links. We would like to go in the other direction, starting with a Legendrian link and constructing a corresponding electromagnetic field for it.

We define the map $\varphi=(\alpha,\beta)|_{t=0}:\mathbb{R}^3\cup\{\infty\}\to S^3$. The particular choice of $(\alpha,\beta)$ in Equation (\ref{eq:stereo3}) does not only determine a contact structure, but also provides us with an explicit orthonormal basis of the plane $\xi_p$ in $T_pS^3$ for all $p\in S^3\backslash\{(1,0)\}$, given by
\begin{equation}
\xi_p=\text{span}\{v_1,v_2\}
\end{equation}


where $v_1$ and $v_2$ are given by
\begin{align}
\label{eq:v1}
v_1&=\varphi_*\left(\frac{(x^2+y^2+z^2+1)^3}{8} \text{Re}\left(\nabla \alpha\bigr\rvert_{t=0} \times \nabla \beta\bigr\rvert_{t=0}\right)\right)\nonumber\\
&=-x_2 \frac{\partial}{\partial x_1}+y_2\frac{\partial}{\partial y_1}+x_1\frac{\partial}{\partial x_2}-y_1\frac{\partial}{\partial y_2},\nonumber\\
v_2&=\varphi_*\left(\frac{(x^2+y^2+z^2+1)^3}{8}\text{Im}\left(\nabla \alpha\bigr\rvert_{t=0} \times \nabla \beta\bigr\rvert_{t=0}\right)\right)\nonumber\\
&=-y_2 \frac{\partial}{\partial x_1}-x_2\frac{\partial}{\partial y_1}+y_1\frac{\partial}{\partial x_2}+x_1\frac{\partial}{\partial y_2}.
\end{align}
They are pushforwards of multiples of $\text{Re}(\nabla \alpha \times \nabla \beta)|_{t=0}$ and $\text{Im}(\nabla \alpha \times \nabla \beta)|_{t=0}$ by $\varphi$. It is easy to see from these expressions that $v_1$ and $v_2$ are orthonormal and span the contact plane $\xi_p$ at each point $p\in S^3\backslash\{(1,0)\}$. The point $p=(1,0)$ is excluded, since it is $(1,0)=\varphi(\infty)$.

A magnetic field $\mathbf{B}$ constructed using Bateman's method satisfies
\begin{align}
\label{eq:Bv1v2}
\mathbf{B}&=\text{Im}(\mathbf{F})\nonumber\\
&=\text{Re}(h(\alpha,\beta))\text{Im}(\nabla \alpha \times \nabla \beta)+\text{Im}(h(\alpha,\beta))\text{Re}(\nabla \alpha \times \nabla \beta),
\end{align}
while the electric field $\mathbf{E}$ satisfies
\begin{align}
\label{eq:Ev1v2}
\mathbf{E}&=\text{Re}(\mathbf{F})\nonumber\\
&=\text{Re}(h(\alpha,\beta))\text{Re}(\nabla \alpha \times \nabla \beta)-\text{Im}(h(\alpha,\beta))\text{Im}(\nabla \alpha \times \nabla \beta).
\end{align}
In particular, both fields are at every point a linear combination of $\text{Re}(\nabla \alpha \times \nabla \beta)$ and $\text{Im}(\nabla \alpha \times \nabla \beta)$ and their pushforwards by $\varphi$ are linear combinations of $v_1$ and $v_2$. The fact that $v_1$ and $v_2$ are a basis for the contact plane $\xi_p$ for all $p\in S^3\backslash\{(1,0)\}$ implies that Equations (\ref{eq:Bv1v2}) and (\ref{eq:Ev1v2}) provide an alternative proof of Lemma \ref{revlemma}. Hence every closed field line must be a Legendrian knot and the holomorphic function $h$ describes the coordinates of the field with respect to this preferred basis.

Suppose now that we have an $n$-component Legendrian link $L=L_1\cup L_2\cup\ldots\cup L_n$ with respect to the standard contact structure on $S^3$, with $(1,0)\not\in L$, a subset $I\subset\{1,2\ldots,n\}$, and a non-zero section $X$ of its tangent bundle $TL\subset \xi_0\subset TS^3$. We can define a complex-valued function $H:L\to\mathbb{C}$ given by
\begin{align}
\label{eq:H}
\text{Re}(H(z_1,z_2))&=X_{(z_1,z_2)}\cdot v_1,&&\text{for all }(z_1,z_2)\in L_i, i\in I\nonumber\\
\text{Im}(H(z_1,z_2))&=-X_{(z_1,z_2)} \cdot v_2,&&\text{for all }(z_1,z_2)\in L_i, i\in I,\nonumber\\
\text{Re}(H(z_1,z_2))&=X_{(z_1,z_2)}\cdot v_2,&&\text{for all }(z_1,z_2)\in L_i, i\notin I\nonumber\\
\text{Im}(H(z_1,z_2))&=X_{(z_1,z_2)} \cdot v_1,&&\text{for all }(z_1,z_2)\in L_i, i\notin I,
\end{align} 
where $\cdot$ denotes the standard scalar product in $\mathbb{R}^4=T_{(z_1,z_2)}\mathbb{C}^2$.

\begin{proposition}
If there is an open neighbourhood $U$ of $S^3\subset\mathbb{C}^2$ and a holomorphic function $h:U\to\mathbb{C}$ with $h|_L=H$, then the corresponding electromagnetic field $\mathbf{F}=h(\alpha, \beta)\nabla\alpha\times\nabla\beta$ at $t=0$ has closed field lines ambient isotopic to (the mirror image of) $L$, with closed electric field lines in the shape of (the mirror image of) $\bigcup_{i\in I}L_i$ and magnetic field lines in the shape of (the mirror image of) $\bigcup_{i\notin I}L_i$.
\end{proposition}
\textit{Proof}: For every point $q\in \varphi^{-1}(\bigcup_{i\notin I}L_i)$ we have
\begin{align}
\label{eq:tang}
\mathbf{B}|_{t=0}(q)=&\left(\text{Re}(h(\alpha,\beta))\text{Im}(\nabla\alpha\times\nabla\beta)\right.\nonumber\\
&\left.+\text{Im}(h(\alpha,\beta))\text{Re}(\nabla\alpha\times\nabla\beta)\right)\Bigr\rvert_{t=0,(x,y,z)=q}\nonumber\\
=&\frac{8}{(|q|^2+1)^3}\left(((\text{Re}(H(\alpha,\beta))(\varphi^{-1})_*(v_2)\right.\nonumber\\
&\left.+\text{Im}(H(\alpha,\beta))(\varphi^{-1})_*(v_1))\right)\Bigr\rvert_{t=0,(x,y,z)=q}\nonumber\\
=&\frac{8}{(|q|^2+1)^3} (\varphi^{-1})_*(X_{(\alpha,\beta)})\Bigr\rvert_{t=0,(x,y,z)=q},
\end{align}
where $|\cdot|$ denotes the Euclidean norm in $\mathbb{R}^3$. The second equality follows from $h|_L=H$ and Equation (\ref{eq:v1}). The last equality follows from the orthonormality of the basis $\{v_1,v_2\}$, the definition of $H$ and the fact that $L$ is Legendrian. Equation (\ref{eq:tang}) states that at $t=0$ the field $\mathbf{B}$ is everywhere tangent to $\varphi^{-1}(\bigcup_{i\notin I}L_i)$. In particular, at $t=0$ the field $\mathbf{B}$ has a set of closed flow lines that is ambient isotopic to the mirror image of $\bigcup_{i\notin I}L_i$ (cf. Remark \ref{remark}).

Similarly, for every $q\in\varphi^{-1}(\bigcup_{i\in I}L_i)$ we have
\begin{align}
\label{eq:tang2}
\mathbf{E}|_{t=0}(q)=&\left(\text{Re}(h(\alpha,\beta))\text{Re}(\nabla\alpha\times\nabla\beta)\right.\nonumber\\
&\left.-\text{Im}(h(\alpha,\beta))\text{Im}(\nabla\alpha\times\nabla\beta)\right)\Bigr\rvert_{t=0,(x,y,z)=q}\nonumber\\
=&\frac{8}{(|q|^2+1)^3}\left(((\text{Re}(H(\alpha,\beta))(\varphi^{-1})_*(v_1)\right.\nonumber\\
&\left.-\text{Im}(H(\alpha,\beta))(\varphi^{-1})_*(v_2))\right)\Bigr\rvert_{t=0,(x,y,z)=q}\nonumber\\
&=\frac{8}{(|q|^2+1)^3} (\varphi^{-1})_*(X_{(\alpha,\beta)})\Bigr\rvert_{t=0,(x,y,z)=q}.
\end{align}
The same arguments as above imply that at $t=0$ the field $\mathbf{E}$ is everywhere tangent to $\varphi^{-1}(\bigcup_{i\in I}L_i)$, so that at $t=0$ the field $\mathbf{E}$ has a set of closed flow lines that is ambient isotopic to $\bigcup_{i\in I}L_i$. \qed

Since the constructed fields are null for all time, the topology of the electric and magnetic field lines does not change, and the fields contain $L$ for all time. We hence have the following corollary.

\begin{corollary}
Let $L=L_1\cup L_2\cup\ldots\cup L_n$ be an $n$-component Legendrian link with respect to the contact structure in $S^3$ with $I\subset \{1,2\ldots,n\}$ and a non-vanishing section of its tangent bundle such that the corresponding function $H:L\to \mathbb{C}$ allows a holomorphic extension $h:U\to\mathbb{C}$ to an open neighbourhood $U$ of $S^3$. Then $\mathbf{F}=h(\alpha,\beta)\nabla\alpha\times\nabla\beta$ has a set of closed field lines that is ambient isotopic to the mirror image of $L$ for all time, with a set of closed electric field lines that is ambient isotopic to the mirror image of $\bigcup_{i\in I}L_i$ for all time and a set of closed magnetic field lines that is ambient isotopic to the mirror image of $\bigcup_{i\notin I}L_i$ for all time. 
\end{corollary}

Therefore, what we have to show in order to prove Theorem \ref{thm:main} is that every link type (with every choice of a subset of its components) has a Legendrian representative as in the corollary. 

\section{The proof of the theorem}
\label{sec:proof}

We have seen in the previous section that Theorem \ref{thm:main} can be proven by showing that every link type has a Legendrian representative for which a certain function has a holomorphic extension. Questions like this, regarding the existence of holomorphic extensions of functions defined on a subset of $\mathbb{C}^m$, are important in the study of complex analysis in $m$ variables and are in general much more challenging when $m>1$. In this section, we first prove that every link type has a Legendrian representative with certain properties regarding real analyticity. We then review a result from complex analysis by Burns and Stout that guarantees that for this class of real analytic submanifolds of $\mathbb{C}^2$ contained in $S^3$ the desired holomorphic extension exists, thereby proving Theorem \ref{thm:main}.

\begin{lemma}
\label{lem}
Every link type has a real analytic Legendrian representative $L$ that admits a non-zero section of its tangent bundle, such that for any given subset $I$ of its set of components the corresponding function $H:L\to\mathbb{C}$ as in Equation (\ref{eq:H}) is real analytic.
\end{lemma}
\textit{Proof}: The lemma is essentially proved in \cite{rudolphtt2}, where it is shown that every link has a Legendrian representative $L$ (with respect to the contact structure in $S^3$) that is the image of a smooth embedding, given by a Laurent polynomial $\eta_i=(\eta_{i,1},\eta_{i,2}):S^1\to S^3\subset\mathbb{C}^2$ in $\rme^{\rmi \chi}$ for each component $L_i$. The set of functions $\eta_i$ in \cite{rudolphtt2} is obtained by approximating some smooth embedding, whose image is a Legendrian link $L'$ of the same link type as $L$. It is a basic exercise in contact topology to show that we can assume that $(1,0)\not\in L'$ \cite{etnyre} and hence also $(1,0)\not\in L$. 

Since each $\eta_i$ is a real analytic embedding, the inverse $\eta_i^{-1}:L\to S^1$ is real analytic in $x_1$, $y_1$, $x_2$ and $y_2$ for all $i=1,2,\ldots,n$. Likewise $\partial_\chi\eta_i:S^1\to TL_i$ is real analytic in $\chi$ and non-vanishing, since $\eta_i$ is an embedding. It follows that the composition $X\defeq(\partial_\chi \eta_i)\circ\eta_i^{-1}:L_i\to TL_i$ is a real analytic non-vanishing section of the tangent bundle of $L_i$ for all $i=1,2,\ldots,n$. Equations (\ref{eq:H}) and (\ref{eq:v1}) then directly imply that $H$ is also real analytic, no matter which subset $I$ of the components of $L$ is chosen. \qed

It was shown in \cite{rudolphtt} that a link $L$ in $S^3$ is a real analytic Legendrian link if and only if it is a \textit{totally tangential} $\mathbb{C}$\textit{-link}, i.e., $L$ arises as the intersection of a complex plane curve and $S^3$ that is tangential at every point. Recall from Remark \ref{remark} that the torus links constructed in \cite{kedia} arise in this way, where the complex plane curve is $z_1^pz_2^q-1=0$ and the radius of the 3-sphere is chosen appropriately. Links that arise as transverse intersections of complex plane curves and the 3-sphere, so-called \textit{transverse $\mathbb{C}$-links} or, equivalently, \textit{quasipositive links}, have been studied as stable vortex knots in null electromagnetic fields in \cite{quasi}.

Following Burns and Stout \cite{burns} we call a real analytic submanifold $\Sigma$ of $\mathbb{C}^2$ that is contained in $S^3$ an \textit{analytic interpolation manifold (relative to the 4-ball $B$)} if every real analytic function $\Sigma\to\mathbb{C}$ is the restriction to $\Sigma$ of a function that is holomorphic on some neighbourhood of $B$. The neighbourhood depends on the function in question.

\begin{theorem}[Burns-Stout \cite{burns}]
\label{thm:burns}
$\Sigma$ is an analytic interpolation manifold if and only if $T_p(\Sigma)\subset T_p^{\mathbb{C}}(S^3)$ for every $p\in\Sigma$, where $T_p^{\mathbb{C}}(S^3)$ denotes the maximal complex subspace of $T_p(S^3)$.
\end{theorem}

The result stated in \cite{burns} holds in fact for more general ambient spaces and their boundaries, namely strictly pseudo-convex domains with smooth boundaries. The open 4-ball $B$ with boundary $\partial B=S^3$ is easily seen to be an example of such a domain.

\textit{Proof of Theorem \ref{thm:main}}: By Lemma \ref{lem} every link type can be represented by a real analytic Legendrian link $L$. It is thus a real analytic submanifold of $\mathbb{C}^2$ that is contained in $S^3$. The condition $T_pL\subset T_p^{\mathbb{C}}(S^3)$ is equivalent to $L$ being a Legendrian link with respect to the standard contact structure on $S^3$. Hence $L$ is an analytic interpolation manifold. Since Lemma \ref{lem} also implies that for every choice of $I$ the function $H:L\to \mathbb{C}$ can be taken to be real analytic, Theorem \ref{thm:burns} implies that $H$ is the restriction of a holomorphic function $h:U\to\mathbb{C}$, where $U$ is some neighbourhood of $S^3$.

The discussion in Section \ref{sec:leg} shows that the electromagnetic field
\begin{equation}
\mathbf{F}=h(\alpha,\beta)\nabla\alpha\times\nabla\beta
\end{equation}
has a set of closed electric field lines in the shape of the mirror image of $\bigcup_{i\in I}L_i$ and a set of closed magnetic field lines in the shape of the mirror image of $\bigcup_{i\notin I}L_i$ at time $t=0$. Since the constructed field is null for all time, $\mathbf{F}$ contains these links for all time, which concludes the proof of Theorem \ref{thm:main}, since every link has a mirror image. \qed

\section{Discussion}
\label{sec:disc}
We showed that every link type arises as a set of stable electric and magnetic field lines in a null electromagnetic field. Since these fields are obtained via Bateman's construction, they share some properties with the torus link fields in \cite{kedia}. They are for example shear-free and have finite energy.

However, since the proof Theorem \ref{thm:main} only asserts the existence of such fields, via the existence of a holomorphic function $h$, other desirable properties of the fields in \cite{kedia} are more difficult to investigate. The electric and magnetic field lines in \cite{kedia} lie on the level sets of $\text{Im}(\alpha^p\beta^q)$ and $\text{Re}(\alpha^p\beta^q)$. At this moment, it is not clear (and doubtful) if the fields in Theorem \ref{thm:main} have a similar integrability property. It is, however, very interesting that the relevant function $z_1^pz_2^q$, whose real/imaginary part is constant on integral curves of the (pushforward of the) magnetic/electric field, is (up to an added constant) exactly the complex plane curve whose totally tangential intersection with $S^3$ gives the $(p,-q)$-torus link. In light of this observation, we might conjecture about the fields in Theorem \ref{thm:main}, which contain $L$, that if the electric/magnetic field lines really lie on the level sets of a pair of real functions, then the real and imaginary parts of $F$ would be natural candidates for such functions, where $F=0$ intersects $S^3$ totally tangentially in the mirror image of $L$. So far $z_1^pz_2^q-1=0$ is the only explicit example of such a function (resulting in the $(p,-q)$-torus link) that the author is aware of, even though it is known to exist for any link. It is this lack of explicit examples and concrete constructions that makes it difficult to investigate this conjecture and other properties of the fields from Theorem \ref{thm:main}.

Kedia et al. also obtained concrete expressions for the helicity of their fields \cite{kedia}. Again, the lack of concrete examples makes it difficult to obtain analogous results.

Since the fields in Theorem \ref{thm:main} are obtained via Bateman's construction, all their Poynting fields at $t=0$ are tangent to the fibers of the Hopf fibration. It is still an open problem to modify the construction, potentially via a different choice of $\alpha$ and $\beta$ to obtain knotted fields, whose underlying Poynting fields give more general Seifert fibrations.




\end{document}